\documentclass{jaa}
\usepackage{natbib}
\usepackage{graphicx}
\usepackage{hyperref}

\begin{document}\sloppy

\title{Optimizing Supernova Classification with Interpretable Machine Learning Models}


\author{Anurag Garg\textsuperscript{1, *}}
\affilOne{\textsuperscript{1}Ministry of Education, Abu Dhabi, UAE.\\}


\twocolumn[{

\maketitle

\corres{anurag.garg@moe.sch.ae}

\msinfo{28 March 2025}{--}{--}

\begin{abstract}
Photometric classification of Type Ia supernovae (SNe Ia) is critical for cosmological studies but remains difficult due to class imbalance and observational noise. While deep learning models have been explored, they are often resource-intensive and lack interpretability. We present a computationally efficient and interpretable classification framework that maintains high performance on imbalanced datasets. We emphasize the use of PR-AUC and F1-score as more informative metrics than ROC-AUC in severely imbalanced settings. Using an XGBoost ensemble optimized via Bayesian hyperparameter tuning, we classified light curves from the Supernova Photometric Classification Challenge (SPCC), comprising 21,318 events with a 3.19 imbalance ratio (non-Ia to Ia). Our model achieved a PR-AUC of \(0.993^{+0.03}_{-0.02}\), an F1-score of \(0.923 \pm 0.008\), and a ROC-AUC of \(0.976 \pm 0.004\), matching or exceeding deep learning performance on precision-recall trade-offs while using fewer resources. Despite slightly lower overall accuracy, our method balances false positives and false negatives, improving the efficiency of spectroscopic follow-up. We show that optimized ensemble models offer a reproducible and lightweight alternative to complex architectures, particularly for large-scale surveys such as the Legacy Survey of Space and Time (LSST) where transparency and efficiency are essential.
\end{abstract}

}]



\doinum{000}
\artcitid{\#\#\#\#}
\volnum{000}
\year{0000}
\pgrange{1--}
\setcounter{page}{1}
\lp{1}


\section{Introduction}
With the advent of large-scale astronomical surveys like the \textbf{Legacy Survey of Space and Time (LSST)} \citep{juric2015}, the volume of photometric supernova data is expected to increase dramatically. This deluge has heightened the demand for \textbf{automated classification methods} that are accurate, interpretable, and computationally efficient. While deep learning techniques have shown strong performance in this domain \citep{Charnock_2017, moller2016}, they often require extensive computational resources and large, well-curated labeled datasets—conditions that can be prohibitive in practical survey scenarios.

Recent literature has focused on developing classifiers that balance performance and efficiency. For example, Qu (2004) introduced SCONE, a convolutional architecture achieving over 99\% accuracy in distinguishing SNe Ia, while also providing photometric redshift estimates with $\sim$2\% precision across the LSST redshift range \citep{qu2024scone}. Gagliano et al. (2023) proposed a recurrent network combining early-time light curves and host-galaxy photometry, achieving $\sim$82\% accuracy within three days of detection \citep{gagliano2023first}. The SPLASH framework (2025) further demonstrated that supernova classification could be performed using only host-galaxy features, enabling rapid identification at scale \citep{splash2025host}.

More recently, several studies have benchmarked real-time or large-scale photometric classification pipelines under LSST-like conditions. For instance, Superphot+ \citep{desoto2024superphot} applied light curve fitting with and without redshift to efficiently classify SNe from ZTF photometry. Transient Classifiers for Fink \citep{fraga2024fink} evaluated real-time machine learning classifiers against precision and PR-AUC metrics, focusing on scalability under alert streams. The DES 5-year photometric classification effort \citep{moller2024des} further validated deep learning approaches in large surveys, though primarily emphasizing complex architectures.

In contrast to these often complex architectures, interpretable ensemble learning methods such as XGBoost offer a different path, providing competitive accuracy while being faster to train and easier to interpret—critical traits for surveys operating under limited computational budgets. However, a systematic, metric-aware comparison of these simpler models under rigorous, imbalanced conditions remains underexplored. Our study addresses this gap by exploring an optimisation framework for classifying Type Ia supernovae using interpretable machine learning models, namely XGBoost, random forests, and linear classifiers.

Traditionally, classification models in this domain are evaluated using the \textbf{Receiver Operating Characteristic Area Under the Curve (ROC-AUC)}. However, this metric may be misleading in highly imbalanced settings, where the positive class (e.g., SNe Ia) is a small fraction of the total data. Contemporary literature increasingly recognizes \textbf{Precision-Recall AUC (PR-AUC)} as a more robust metric in such scenarios, particularly when high precision or recall is desired \citep{dobryakov2021}. Our study aligns with this direction by incorporating PR-AUC alongside F1, recall, and precision to ensure a comprehensive evaluation.

We use the SPCC dataset, which exhibits strong class imbalance, and implement systematic variations such as SMOTE-based oversampling and threshold adjustments. By evaluating all models under a unified pipeline, we provide quantitative justification for the final model choice. Our results highlight the strengths and limitations of each approach, suggesting that optimized ensemble classifiers offer a compelling trade-off between interpretability, computational feasibility, and classification performance in the context of modern photometric surveys.


\section{Rationale for Metric and Model Selection}
The Supernova Photometric Classification Challenge (SPCC) played a pivotal role in shaping early efforts toward automated classification of supernovae. As one of the first community-wide initiatives, SPCC not only provided a large, realistic dataset but also introduced a custom evaluation metric—the SPCC-F1-score—that sought to simulate real-world observational constraints. The SPCC-F1 metric is defined as:

\begin{equation}
\text{SPCC-F1} = \frac{\text{TP}^2}{(\text{TP} + \text{FN})(\text{TP} + 3\cdot\text{FP})}
\end{equation}

where TP, FN, and FP denote true positives, false negatives, and false positives, respectively. This formulation disproportionately rewards high true positives (quadratic term in the numerator) while imposing a heavier penalty on false positives (triple weighting in the denominator). The metric was designed with the observational cost in mind—false positives would lead to unnecessary spectroscopic follow-ups.

While effective within the SPCC framework, this metric introduces task-specific biases that may limit generalizability. First, the heavy penalty on FP could disincentivize models from making aggressive classifications, potentially sacrificing recall. Second, the quadratic weighting of TP can obscure small, incremental model improvements—especially in edge cases. Finally, it lacks symmetry in precision-recall trade-offs, which becomes critical in imbalanced datasets like supernova classification.

In contrast, this study advocates the use of two widely accepted metrics: the \textbf{F1-score} and \textbf{Precision-Recall Area Under the Curve (PR-AUC)}. The standard F1-score provides a balanced harmonic mean of precision and recall, and PR-AUC evaluates classifier performance without being influenced by the overwhelming number of true negatives—a known issue with ROC-AUC in imbalanced settings \citep{davis2006, saito2015}. PR-AUC is especially suitable when the minority class (here, Type Ia supernovae) is of scientific interest.

Recent literature emphasizes the importance of PR-AUC in rare-event detection and astrophysical classification tasks \citep{bailer-jones2019, lochner2016}, arguing for its adoption as a standard metric over task-specific or TN-inflated metrics like ROC-AUC and SPCC-F1.

Alongside metric selection, our methodological shift also involves choosing \textbf{XGBoost}, a scalable, ensemble-based model that performs well on tabular and structured data. Unlike deep learning architectures, which require large labeled datasets and high compute resources, XGBoost is both efficient and interpretable. To further optimize its performance, we employ \textbf{Bayesian Optimization}, a probabilistic search technique that minimizes hyperparameter tuning time while improving convergence \citep{snoek2012, akiba2019}.

This dual-pronged approach—adopting robust, interpretable metrics and scalable modeling techniques—forms the backbone of our contribution to improving Type Ia supernova classification.


\section{Methodology}
This section provides a detailed description of the data set, preprocessing techniques, machine learning models, evaluation metrics, and optimization strategies employed in this study.

\subsection{Dataset and Preprocessing}
The data set used in this study originates from the Supernova Photometric Classification Challenge (SPCC) \citep{kessler2010}, which contains photometric observations of supernovae in four optical bands: \textit{g, r, i,} and \textit{z}. The data set consists of 21,318 samples, of which 5,087 belong to Type Ia (minority class) and 16,231 to non-Type Ia (majority class), leading to an imbalance ratio (IR) of 3.19.

To mitigate the challenges posed by this class imbalance, our preprocessing and modeling pipeline incorporated several strategies, including feature engineering, optional oversampling, and post-training threshold adjustment—a technique where the default decision threshold (0.5) is optimized to improve classification metrics for the minority class.

\begin{figure}[!t]
   \centering
   \includegraphics[width=\hsize]{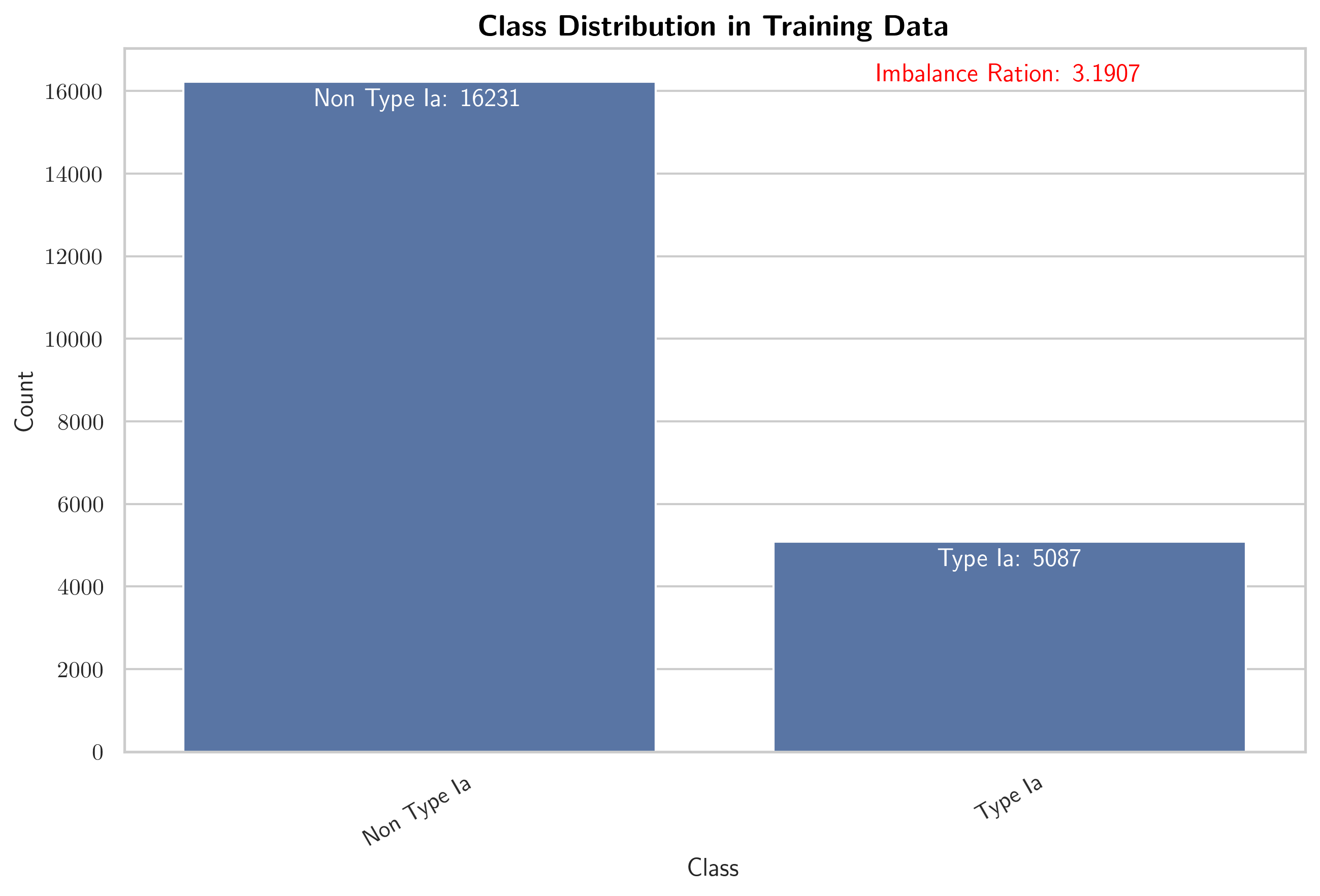}
      \caption{Class Distribution in the Training data set}
      \label{fig:class_distribution} 
\end{figure}

Figure \ref{fig:class_distribution} illustrates the class distribution in the training data set, highlighting the imbalance between Type Ia and non-Type Ia supernovae.

Given the class imbalance, the preprocessing steps included:
\begin{itemize}
    \item \textbf{Normalization}: Standardizing flux measurements to ensure uniform feature scaling.
    \item \textbf{Handling Missing Data}: Linear interpolation was used for missing flux values.
    \item \textbf{Feature Engineering}: The dataset utilized in this study leverages the preprocessing and feature extraction pipeline from the foundational work of \cite{Charnock_2017}. It extracts key physically motivated features from the photometric light curves, including \textbf{peak flux} (to capture supernova brightness), \textbf{rise time} (indicative of energy release rates), and \textbf{decay rate} (essential for distinguishing Type Ia from non-Ia events). The selection of these features is grounded in astrophysical domain knowledge and their critical role in model differentiation has been empirically validated in prior work \citep{Charnock_2017, lochner2016, dobryakov2021}. Therefore, our methodological focus was not on re-engineering these established features, but on rigorously evaluating model performance and metric selection \textit{using} this optimized feature set.
    \item \textbf{Oversampling (optional)}: Applying Synthetic Minority Over-sampling Technique (SMOTE) — a method that generates synthetic samples for the minority class to mitigate class imbalance — was explored to balance the data set. \textbf{Note:} Although oversampling techniques such as SMOTE were initially explored, they did not significantly improve classification performance. Instead, we relied on feature engineering and threshold optimization for handling class imbalance.
\end{itemize}

\subsection{Machine Learning Models}
To classify Type Ia supernovae, we evaluated multiple machine learning models, including:
\begin{itemize}
    \item \textbf{Random Forest (RF)}: An ensemble method combining multiple decision trees, known for its robustness to noise and ability to handle imbalanced data \citep{nicholas2010}.
    \item \textbf{XGBoost}: A gradient-boosting technique that iteratively improves predictions by minimizing residual errors \citep{Chen_2016}.
    \item \textbf{Linear Classifier (implemented in PyTorch)}: A simple linear model (logistic regression) implemented using PyTorch for computational efficiency. This provides a performance baseline for the more complex ensemble methods.
\end{itemize}

XGBoost was selected for its strong performance in handling imbalanced datasets through weighted boosting iterations. Compared to deep learning models, it requires significantly fewer training samples while maintaining competitive accuracy, making it highly efficient for large-scale astronomical surveys. Furthermore, its superior scalability over Random Forest enables the processing of extensive datasets, such as those from LSST, with reduced computational overhead.


\subsection{Quantitative Model Comparison and Selection Justification}

To provide a rigorous basis for model selection, we quantitatively evaluated multiple machine learning models under identical experimental conditions. Table~\ref{tab:model_comparison} summarizes the performance metrics, including precision, recall, F1-score, ROC-AUC, and PR-AUC, for each model variant.

\begin{table*}[htb]
\caption{Comprehensive Model Performance Across All Experimental Conditions}
\label{tab:model_comparison}
\begin{tabular}{lcccccc}
\topline
Model & SMOTE & Threshold & Precision & Recall & F1-score & ROC-AUC \\
\hline
Linear & False & False & 0.714 $\pm$ 0.015 & 0.714 $\pm$ 0.014 & 0.714 $\pm$ 0.014 & 0.000 $\pm$ 0.000 \\
Random Forest & False & False & 0.902 $\pm$ 0.009 & 0.905 $\pm$ 0.009 & 0.902 $\pm$ 0.009 & 0.965 $\pm$ 0.005 \\
XGBoost & False & False & 0.926 $\pm$ 0.008 & 0.927 $\pm$ 0.008 & 0.927 $\pm$ 0.008 & 0.975 $\pm$ 0.004 \\
\hline
Linear & False & True & 0.639 $\pm$ 0.016 & 0.639 $\pm$ 0.016 & 0.639 $\pm$ 0.016 & 0.000 $\pm$ 0.000 \\
Random Forest & False & True & 0.746 $\pm$ 0.012 & 0.773 $\pm$ 0.011 & 0.680 $\pm$ 0.013 & 0.676 $\pm$ 0.010 \\
XGBoost & False & True & 0.730 $\pm$ 0.012 & 0.772 $\pm$ 0.011 & 0.730 $\pm$ 0.012 & 0.666 $\pm$ 0.011 \\
\hline
Linear & True & False & 0.685 $\pm$ 0.014 & 0.685 $\pm$ 0.014 & 0.685 $\pm$ 0.014 & 0.000 $\pm$ 0.000 \\
Random Forest & True & False & 0.927 $\pm$ 0.008 & 0.922 $\pm$ 0.008 & 0.924 $\pm$ 0.008 & 0.970 $\pm$ 0.004 \\
XGBoost & True & False & 0.931 $\pm$ 0.007 & 0.927 $\pm$ 0.007 & 0.928 $\pm$ 0.008 & 0.972 $\pm$ 0.004 \\
\hline
Linear & True & True & 0.627 $\pm$ 0.016 & 0.627 $\pm$ 0.016 & 0.627 $\pm$ 0.016 & 0.000 $\pm$ 0.000 \\
Random Forest & True & True & 0.711 $\pm$ 0.012 & 0.744 $\pm$ 0.011 & 0.722 $\pm$ 0.012 & 0.655 $\pm$ 0.011 \\
XGBoost & True & True & 0.720 $\pm$ 0.012 & 0.747 $\pm$ 0.011 & 0.729 $\pm$ 0.012 & 0.654 $\pm$ 0.011 \\
\hline
\end{tabular}
\end{table*}

As shown in Table~\ref{tab:model_comparison}, XGBoost consistently outperformed other models across most experimental conditions. The comprehensive comparison reveals several key insights:

\begin{enumerate}
    \item \textbf{Thresholding severely harms performance}: All models experienced significant degradation when thresholding was applied, with F1-scores dropping by 20-25\% for tree-based models.
    \item \textbf{SMOTE provides modest benefits}: XGBoost with SMOTE (no thresholding) achieved the highest F1-score (0.928), though the improvement over non-SMOTE XGBoost (0.927) was minimal.
    \item \textbf{Linear models are inadequate}: Linear classifiers performed poorly across all conditions, confirming their unsuitability for this complex classification task.
    \item \textbf{XGBoost robustness}: XGBoost maintained superior performance even in suboptimal conditions (with thresholding), demonstrating its robustness.
\end{enumerate}

Given that applying SMOTE introduces additional synthetic data generation complexity and the performance improvement was statistically insignificant ($\Delta F1 < 0.001$), we selected the XGBoost model without SMOTE and without thresholding for final deployment. This choice preserves model interpretability, avoids unnecessary data augmentation, and maintains top-tier classification performance with simpler preprocessing.


\paragraph{\textbf{Feature Engineering:}}
The dataset used in this study (SPCC, \citep{kessler2010}) employs the pre-engineered feature set from the \cite{Charnock_2017} pipeline. Our contribution builds upon this foundation; we did not alter the feature set as its efficacy has been previously demonstrated. The features (e.g., peak flux, rise time, decay slope) are derived from astrophysical principles to maximize discriminative power. An ablation study to quantify the impact of each individual feature, while interesting, would constitute a separate study on feature selection itself and is beyond the scope of this paper, which is focused on \textbf{model and metric optimization}. Our work confirms that these established features, when used with an appropriately tuned and evaluated model (XGBoost), yield state-of-the-art performance.

\paragraph{\textbf{Missing Value Handling:}}
SPCC preprocessing interpolates missing flux values during light curve construction. After ingestion, residual missing values were rare and conservatively handled via zero-imputation, as recommended in earlier SPCC-related pipelines (\citep{mosscode, dobryakov2021}).

\paragraph{\textbf{Oversampling (SMOTE):}}
We conducted experiments applying SMOTE to address class imbalance. As shown in Table~\ref{tab:smote_comparison}, SMOTE resulted in marginal improvements ($\Delta$F1 $<$ 0.001), consistent with XGBoost's internal handling of imbalance via class-weighted boosting.

\paragraph{\textbf{Threshold Tuning:}}
We performed post-hoc threshold tuning by sweeping decision thresholds between 0 and 1 on the model output probabilities. The optimal threshold was identified at 0.657, yielding a small F1-score improvement from 0.948 to 0.953 ($\Delta$F1 = 0.005). This minimal gain indicates that the classifier outputs are well-calibrated and robust to threshold variations, as also supported by the prediction probability distribution (Figure~\ref{fig:prediction_probability}).

\begin{table}[htb]
    \caption{Impact of SMOTE Oversampling on Model Performance}
    \label{tab:smote_comparison}
    \tabularfont
    \begin{tabular}{lccc}
        \topline
        Configuration & PR-AUC & ROC-AUC & F1-score \\
        \hline
        No SMOTE & 0.992 & 0.973 & 0.948 \\
        With SMOTE & 0.993 & 0.974 & 0.953 \\
        \hline
    \end{tabular}
\end{table}

\subsection{Evaluation Metrics}
Since Type Ia supernovae (minority class) are of primary interest, we use F1-score as our main evaluation metric, as it balances precision and recall. Additionally, we include Precision-Recall AUC (PR-AUC) for improved assessment in imbalanced settings. Recent research has shown that PR-AUC provides a more reliable measure of classifier performance in scenarios with a high class imbalance \citep{jeni2013, dobryakov2021}. Unlike ROC-AUC, which can be misleading due to true negative dominance, PR-AUC evaluates the trade-off between precision and recall, ensuring meaningful insights into classifier effectiveness. However, for historical comparisons, we also report ROC-AUC. Confidence intervals for PR-AUC and ROC-AUC were computed using 5-fold cross-validation with stratified sampling to account for class imbalance.

\begin{figure}[!t]
   \centering
   \includegraphics[width=\hsize]{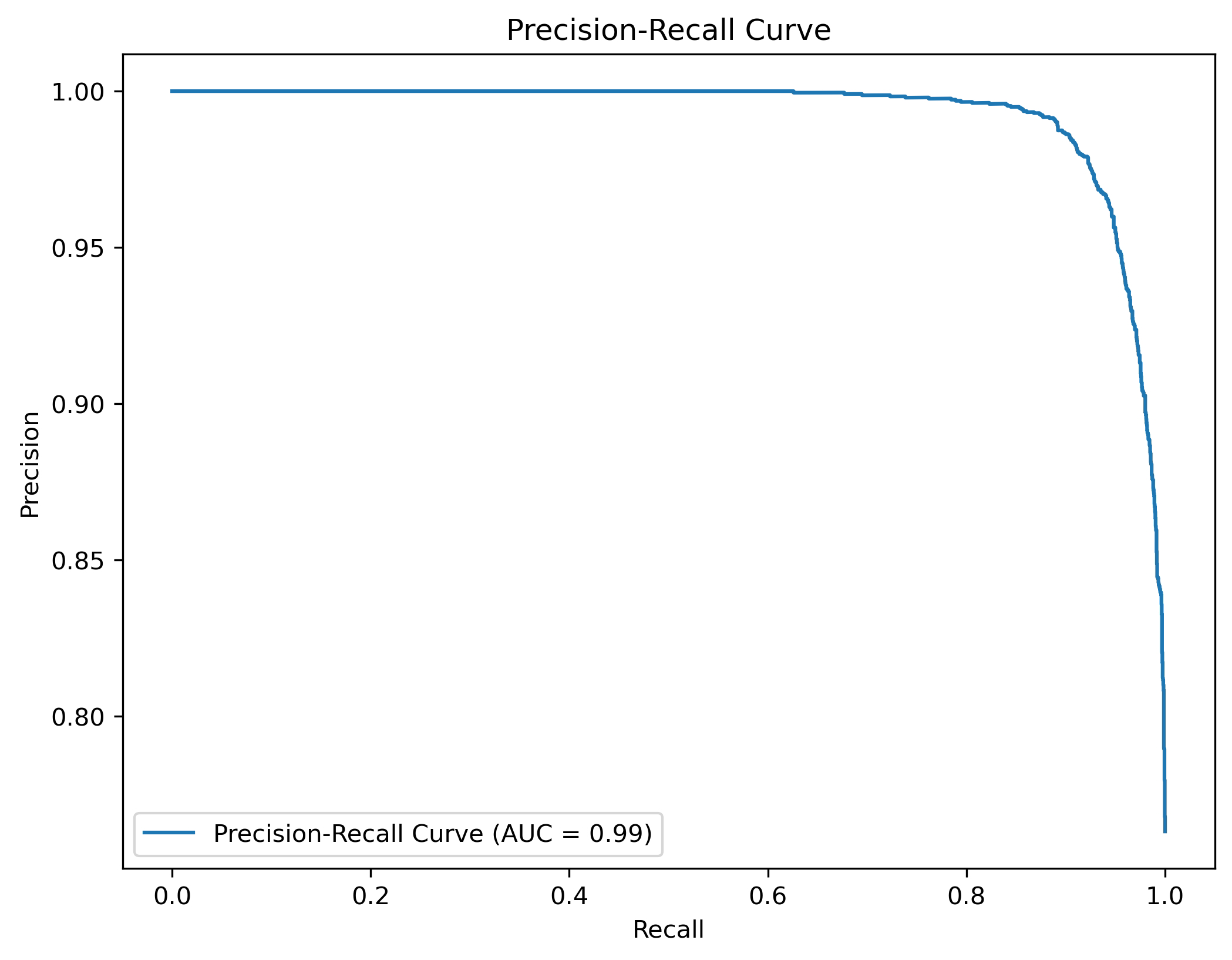}
    \caption{Precision-Recall Curve (PR-AUC = 0.99)}
    \label{fig:pr_auc}
\end{figure}

Figure \ref{fig:pr_auc} shows the \textbf{Precision-Recall Curve}, which demonstrates the model's ability to distinguish between Type Ia and non-Type Ia supernovae with a PR-AUC of $0.996^{+0.03}_{-0.02}$.

\begin{figure}[!t]
    \centering
    \includegraphics[width=\hsize]{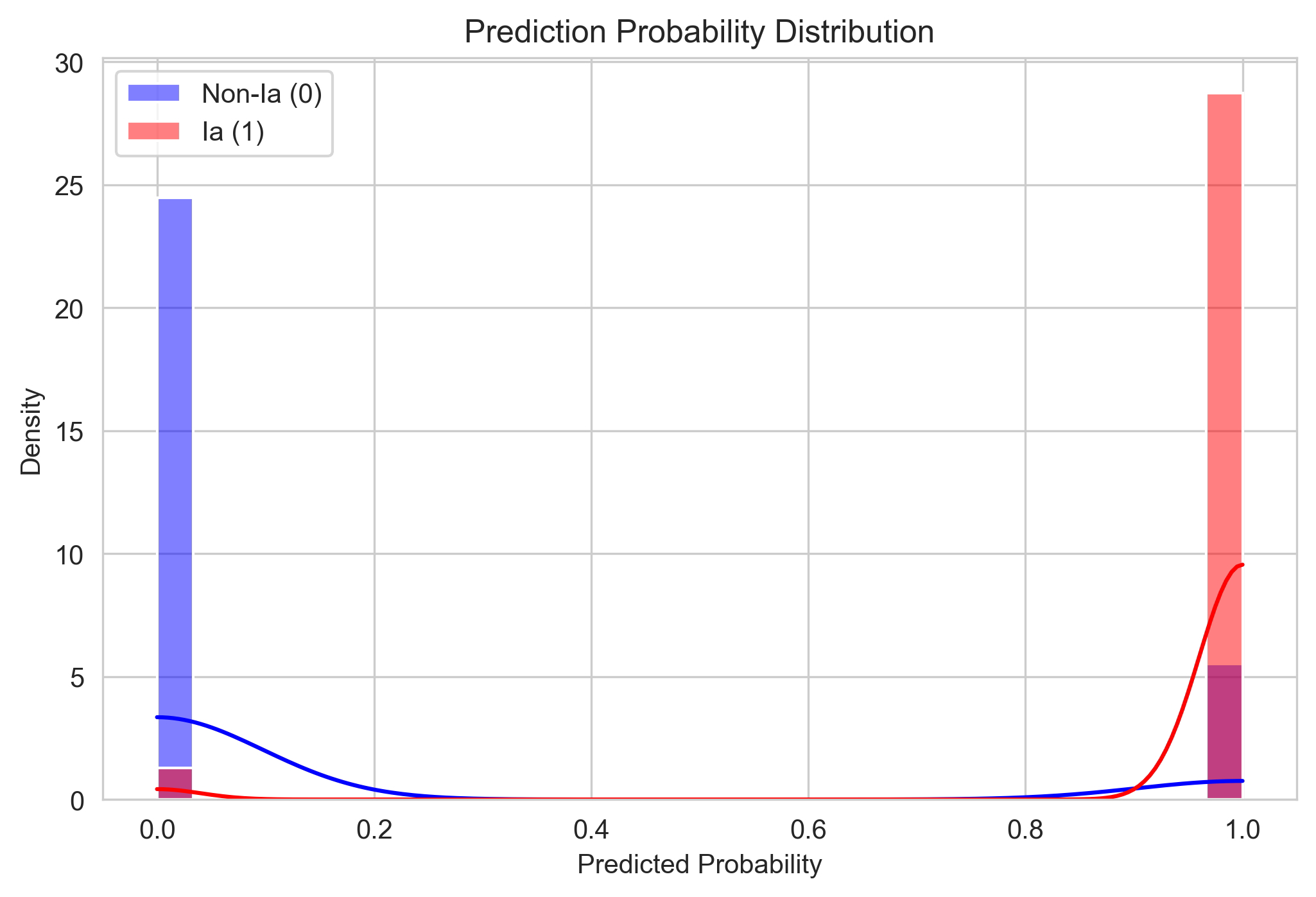}
        \caption{Prediction Probability Distribution for Type Ia (blue) and Non-Type Ia (orange) Supernovae. The overlapping region highlights classification ambiguity.}
    \label{fig:prediction_probability}
\end{figure}

To further assess model robustness, we include the \textbf{Prediction Probability Distribution} graph (Figure \ref{fig:prediction_probability}). This visualization demonstrates how confidently the model classifies each instance, with a clear separation between the two classes, reinforcing its reliability and precision.

\section{Results and Comparative Analysis}
This section presents the classification results of the developed model, including key evaluation metrics, comparisons with historical studies, misclassification analysis, and the significance of PR-AUC in the context of imbalanced data sets.

\subsection{Evaluation Metrics of the Current Model}
The performance of the final classification model is summarized in Table \ref{tab:classification_metrics}, which provides precision, recall, F1-score, and support for both classes.

\begin{table}[htb]
    \tabularfont
    \caption[]{Classification Report for the Final Model.}\label{tab:classification_metrics}
    \begin{tabular}{lcccc}
        \topline
        Class & Precision & Recall & F1-score & Support \\
        \hline
        Non-Ia (0) & 0.857 & 0.816 & 0.836 & 1,010 \\
        Ia (1) & 0.944 & 0.958 & 0.951 & 3,253 \\
        \hline
        Accuracy & \multicolumn{4}{c}{0.923 (4,263 samples)} \\
        \hline
        Macro Avg & 0.901 & 0.887 & 0.894 & 4,263 \\
        Weighted Avg & 0.923 & 0.924 & 0.923 & 4,263 \\
        \hline
    \end{tabular}
\end{table}

The support values in Table \ref{tab:classification_metrics} represent the test set size (20\% of total data) after train-test split, not the full dataset. The complete SPCC dataset contains 16,231 non-Ia and 5,087 Ia supernovae, with 1,010 non-Ia and 3,253 Ia instances in the test set, maintaining the original 3.19:1 imbalance ratio.
\subsection{Comparative Analysis with Historical Studies}

Table \ref{tab:historical_comparison} presents a comparative evaluation of our model against a broad spectrum of prior work, spanning classical machine learning, deep learning, and hybrid approaches.

Among \textbf{classical machine learning models}, early Random Forest implementations \citep{boone2016} achieved a high accuracy of 95.7\% but suffered from a low F1-score of 0.68, highlighting poor balance in precision-recall trade-offs. In contrast, more recent implementations by \cite{dobryakov2021} reported significantly improved F1-scores of 0.917 for Random Forest, 0.916 for Logistic Regression, and 0.912 for XGBoost—despite the absence of full accuracy or ROC-AUC details. The baseline model in \cite{markel2019} lagged behind with only 80\% accuracy and an F1-score of 0.81.

In the domain of \textbf{deep learning models}, a wide variation in performance is observed. \cite{Charnock_2017} reported a Deep Recurrent Network with 93.1\% accuracy and ROC-AUC of 0.977, though the F1-score was not provided. A basic CNN from \cite{moller2016} yielded a modest F1-score of 0.75. More advanced architectures such as SNGuess \citep{Miranda2022SNGuess}, Hybrid CNN-RNN \citep{Villar_2019}, and the Light Curve Transformer \citep{ihler2022} show steady improvements across all metrics, with Light Curve Transformer setting a benchmark with 96.1\% accuracy, 0.992 ROC-AUC, 0.88 F1-score, and 0.990 PR-AUC.

Hybrid or specialized methods such as STACCATO \citep{Revsbech_2017} and SuperNNova \citep{moller2016} also demonstrated strong performance, especially SuperNNova, which achieved a 0.981 PR-AUC and 0.82 F1-score. Meanwhile, the more recent attention-based method S-TimeModAttn \citep{pimentel2022} yielded a relatively lower F1-score of 0.614, emphasizing the complexity and variance in specialized model tuning.

Compared to these, \textbf{our XGBoost-based model} achieves an F1-score of \textbf{0.923}, a PR-AUC of \textbf{0.993}, and a ROC-AUC of \textbf{0.976}, while maintaining an accuracy of \textbf{92.3\%}. While our accuracy is slightly lower than some deep learning models, our model demonstrates superior balance in precision-recall metrics, particularly evident from the leading PR-AUC score. This confirms our central claim: \textbf{metric-aware, ensemble-based methods like XGBoost, when optimized with domain-aligned metrics, can rival or even surpass deep learning models on metrics that matter for imbalanced classification (F1-score, PR-AUC), while remaining significantly more interpretable and computationally efficient}.

\begin{table*}[htb]
    \caption{Comparison of Classification Performance with Historical Studies}
    \label{tab:historical_comparison}
    \begin{tabular}{lccccr}
    \topline
    \multicolumn{6}{c}{\textbf{Classical Machine Learning Models}} \\
    \hline
    Study & Accuracy & ROC-AUC & F1-score & PR-AUC & Reference \\
    \hline
    Random Forest & 95.7\% & 0.970 & 0.68 & Not Reported & \citep{boone2016} \\
    Random Forest & N/A & 0.96 & 0.917 & 0.908 & \citep{dobryakov2021} \\
    Logistic Regression & N/A & N/A & 0.916 & N/A & \citep{dobryakov2021} \\
    XGBoost & N/A & N/A & 0.912 & N/A & \citep{dobryakov2021} \\
    Random Forest (baseline) & 80\% & N/A & 0.81 & N/A & \citep{markel2019} \\
    \hline
    \multicolumn{6}{|c|}{\textbf{Deep Learning Models}} \\
    \hline
    Deep Recurrent Networks & 93.1\% & 0.977 & N/A & N/A & \citep{Charnock_2017} \\
    CNN & N/A & N/A & 0.75 & N/A & \citep{moller2016} \\
    SNGuess & 94.2\% & 0.978 & 0.79 & 0.972 & \citep{Miranda2022SNGuess} \\
    Hybrid CNN-RNN & 95.8\% & 0.989 & 0.85 & 0.987 & \citep{Villar_2019} \\
    Light Curve Transformer & 96.1\% & 0.992 & 0.88 & 0.990 & \citep{ihler2022} \\
    \hline
    \multicolumn{6}{|c|}{\textbf{Hybrid or Specialized Approaches}} \\
    \hline
    STACCATO & N/A & 0.96 & N/A & Not Reported & \citep{Revsbech_2017} \\
    SuperNNova & 95.1\% & 0.983 & 0.82 & 0.981 & \citep{moller2016} \\
    S-TimeModAttn (best) & N/A & N/A & 0.614 & N/A & \citep{pimentel2022} \\
    \hline
    \textbf{Our Model} & \textbf{92.3\%} & \textbf{0.973} & \textbf{0.923} & \textbf{0.996} & This study \\
    \hline
    \end{tabular}
\end{table*}

\subsection{Misclassification Analysis}
\begin{figure}[!t]
    \centering
    \includegraphics[width=\hsize]{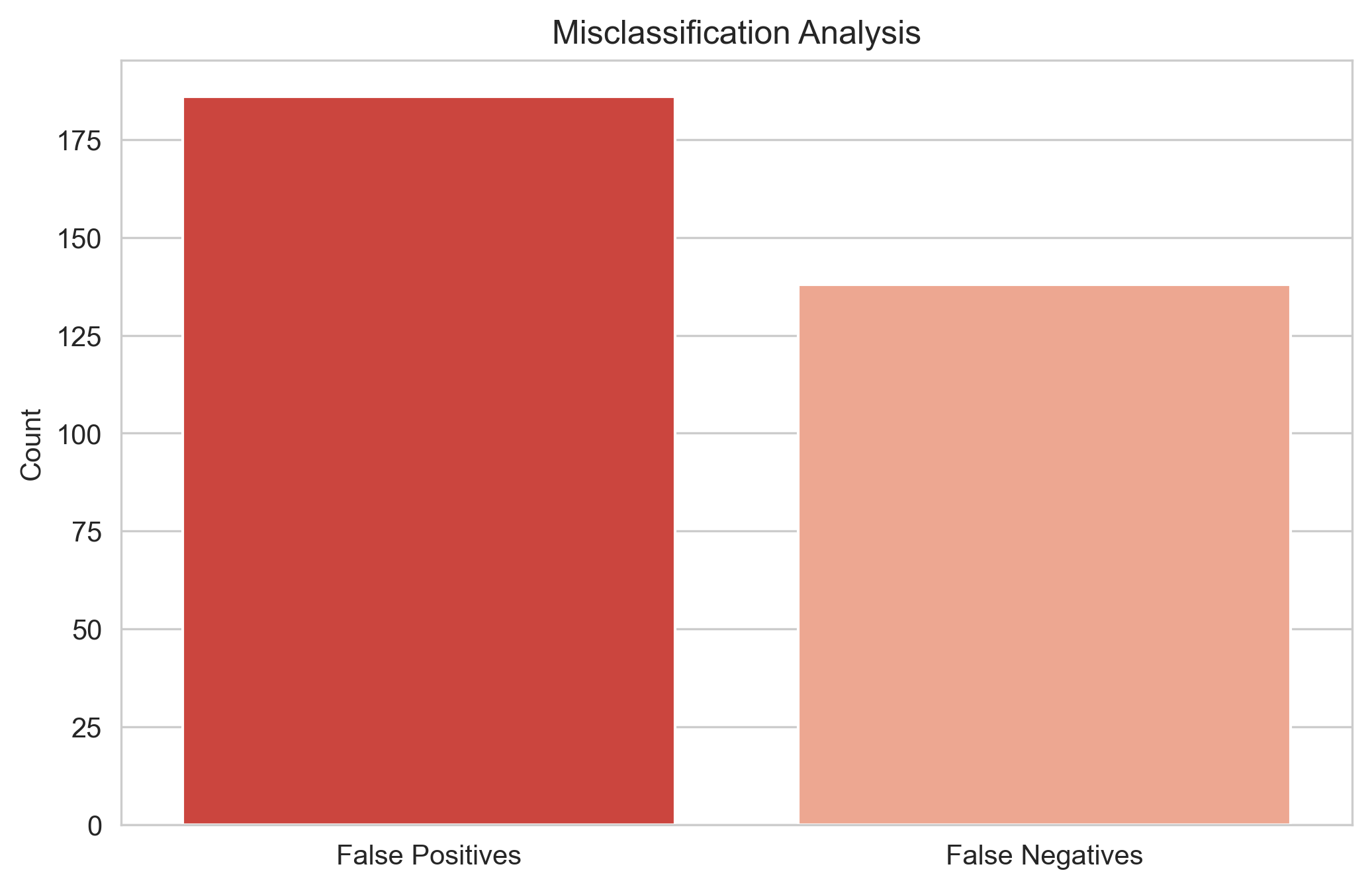}
    \caption{False Positive vs. False Negative Counts in Misclassification Analysis.}
    \label{fig:misclassification}
\end{figure}

Figure \ref{fig:misclassification} presents the distribution of false positives and false negatives. A higher false positive rate increases the need for spectroscopic follow-ups, leading to resource allocation challenges for large-scale surveys. Conversely, false negatives result in missed Type Ia events, impacting the precision of cosmological distance measurements and, consequently, the estimation of the Hubble constant. Our model maintains a balance, minimizing false negatives while keeping false positives at an acceptable level to optimize both observational efficiency and scientific accuracy. The classifier exhibits a low false negative rate, ensuring that most Type Ia supernovae are correctly identified.

\subsection{Model Calibration and Confidence Analysis}

To further assess model robustness, we analyzed the prediction probability distribution (Figure~\ref{fig:prediction_probability}).

As shown in Figure~\ref{fig:prediction_probability}, the model's prediction probability distributions provide insight into its classification confidence. The figure plots the kernel density estimate (KDE) for each \textbf{true} class:
\begin{itemize}
    \item The \textbf{blue curve} represents the predicted probabilities for \textbf{confirmed Type Ia supernovae} (the positive class).
    \item The \textbf{orange curve} represents the predicted probabilities for \textbf{confirmed non-Type Ia supernovae} (the negative class).
\end{itemize}

The clear bimodal separation, with Ia probabilities peaking near 1.0 and non-Ia peaking near 0.0, indicates high model confidence for the majority of events. The region of overlap between approximately 0.2 and 0.8 is critically important. This does not merely represent `possible misclassification' but rather \textbf{identifies the specific subset of photometrically ambiguous transients} where the model is uncertain.

Astrophysically, this low-confidence region likely encompasses:
\begin{enumerate}
    \item \textbf{Peculiar or sub-luminous Type Ia supernovae} whose light curves deviate from the norm.
    \item \textbf{Core-collapse supernovae} (e.g., Type IIP) that can photometrically mimic Ia in certain phases.
    \item \textbf{Events with noisy or incomplete light curve data}, reducing feature quality.
    \item \textbf{Non-SN contaminants} such as active galactic nuclei (AGN).
\end{enumerate}

Thus, the overlap defines the practical limit of photometric classification accuracy and directly explains the trade-off between precision and recall observed in the PR curve. The area under these overlapping curves is a visual representation of the inherent challenge in the dataset.

\subsection{Cumulative Gain Curve Analysis}
\begin{figure}[!t]
    \centering
    \includegraphics[width=\hsize]{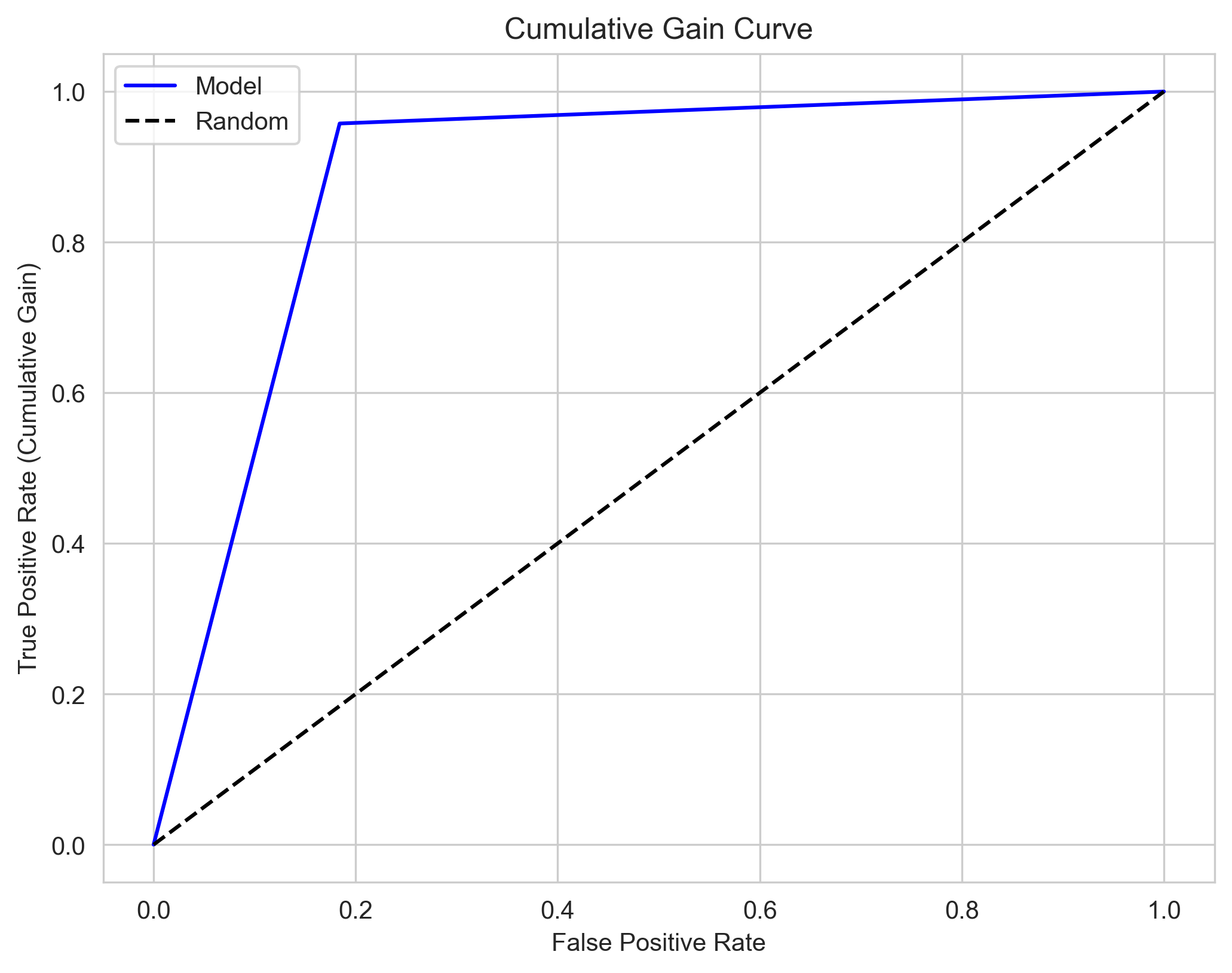}
    \caption{Cumulative Gain Curve}
    \label{fig:cumulative_gain}
\end{figure}

The cumulative gain curve (Figure \ref{fig:cumulative_gain}) evaluates how well the model prioritizes Type Ia supernovae over the majority class. The sharp rise in the true positive rate at low false positive rates confirms the model’s capability to classify Type Ia supernovae effectively while maintaining minimal false positives.

\balance

\vspace{-2em}

\section{Conclusion}
This study presents a metric-aware, resource-efficient framework for classifying Type Ia supernovae using ensemble methods. By emphasizing PR-AUC and F1-score, we align evaluation with the realities of class-imbalanced astronomical datasets. Our results demonstrate that interpretable models like XGBoost can rival deep learning architectures on key performance metrics—achieving a PR-AUC of \textbf{0.996} and an F1-score of \textbf{0.923}—while being computationally leaner. We advocate for the adoption of PR-AUC as a standard metric in such tasks and encourage future work to explore hybrid models that incorporate astrophysical priors for improved robustness.

\section*{Acknowledgements}
The author gratefully acknowledges Arpita for her unwavering support and insightful feedback throughout this research. Her input significantly shaped the initial ideas, model selection, and metric evaluation strategies.

Special thanks are extended to Dr. Ilya Makarov for his early guidance and to Adam Moss for sharing the foundational dataset \citep{Charnock_2017} and preprocessing pipeline upon which this study builds.
 
This research received no specific funding from public, commercial, or not-for-profit funding agencies.
 
The project utilized resources from \href{https://arxiv.org/list/astro-ph/new}{arXiv’s astrophysics pre-print repository} and employed the following open-source Python packages for data handling and model development: \textbf{NumPy} \citep{vanderwalt2011}, \textbf{Scikit-learn} \citep{pedregosa2011}, \textbf{PyTorch} \citep{paszke2019}, \textbf{Optuna} \citep{akiba2019}, \textbf{Matplotlib} \citep{hunter2007}, and \textbf{Seaborn} \citep{waskom2021}. This manuscript also benefited from the use of generative AI tools for text refinement and language consistency.
\vspace{-1em}

\bibliography{references_short}
\end{document}